%% file: main.tex
\documentclass{article}

\usepackage{arxiv}

\usepackage[utf8]{inputenc} % allow utf-8 input
\usepackage[T1]{fontenc}    % use 8-bit T1 fonts
\usepackage{hyperref}       % hyperlinks
\usepackage{url}            % simple URL typesetting
\usepackage{booktabs}       % professional-quality tables
\usepackage{amsfonts}       % blackboard math symbols
\usepackage{amsmath}        % add by fei
\usepackage{bbm}            % add by fei
\usepackage{nicefrac}       % compact symbols for 1/2, etc.
\usepackage{microtype}      % microtypography
\usepackage{cleveref}       % smart cross-referencing
\usepackage{lipsum}         % Can be removed after putting your text content
\usepackage{graphicx}
\usepackage{doi}
\usepackage{subcaption}

\usepackage{derivative}

\title{Recurrent Dynamic Message Passing with Loops for Epidemics on Networks}

\date{}

\author{
    Fei Gao\textsuperscript{1,2,}\thanks{Corresponding author.} ,\quad Jing Liu\textsuperscript{3},\quad Yaqian Zhao\textsuperscript{1,2}
    \\
    \textsuperscript{1}Inspur (Beijing) Electronic Information Industry Co., Ltd, Beijing 100085, China
    \\
    \textsuperscript{2}Inspur Electronic Information Industry Co., Ltd, Jinan 250101, China
    \\
    \textsuperscript{3}CAS Key Laboratory for Theoretical Physics,
    Institute of Theoretical Physics, 
    \\
    Chinese Academy of Sciences, Beijing 100190, China
    \\
    \texttt{feigao.sc@gmail.com},\quad \texttt{jing.liu@itp.ac.cn},\quad \texttt{zhaoyaqian@ieisystem.com}
}

\renewcommand{\headeright}{}
\renewcommand{\undertitle}{}
\renewcommand{\shorttitle}{Recurrent Dynamic Message Passing with Loops for Epidemics on Networks}

\begin{document}
\maketitle

% abstract
\begin{abstract}
    \input{texs/abs.tex}
\end{abstract}


\input{texs/sec_intro.tex}
\input{texs/sec_othermethods.tex}
\input{texs/sec_echochamber.tex}

\input{texs/sec_lrdmp.tex}
\input{texs/sec_experiments.tex}
\input{texs/sec_conclu.tex}

\input{texs/dec.tex}

\bibliographystyle{plain} % 选择合适的样式，例如 "plainnat"
\bibliography{references} % 注意，这里没有文件扩展名 ".bib"
\end{document}

%% file: texs/abs.tex
Several theoretical methods have been developed to approximate prevalence and threshold of epidemics on networks. Among them, the recurrent dynamic message-passing (rDMP) theory offers a state-of-the-art performance by preventing the echo chamber effect in network edges. However, the rDMP theory was derived in an intuitive \textit{ad-hoc} way, lacking a solid theoretical foundation and resulting in a probabilistic inconsistency flaw. Furthermore, real-world networks are clustered and full of local loops like triangles, whereas rDMP is based on the assumption of a locally tree-like network structure, which makes rDMP potentially inefficient on real applications. In this work, for the recurrent-state epidemics, we first demonstrate that the echo chamber effect exits not only in edges but also in local loops, which rDMP-like method can not avoid. We then correct the deficiency of rDMP in a principled manner, leading to the natural introduction of new \textit{higher-order} dynamic messages, extending rDMP to handle local loops. By linearizing the extended message-passing equations, a new epidemic threshold estimation is given by the inverse of the leading eigenvalue of a matrix named \textit{triangular non-backtracking} matrix. 
Numerical experiments conducted on synthetic and real-world networks to evaluate our method, the efficacy of which is validated in epidemic prevalence and threshold prediction tasks. 
% In addition, we discuss the potential benefit of our method in speeding up the solution of the immunization, influence maximization, and robustness optimization problems in the networks.
In addition, our method has the potential to speed up the solution of the immunization, influence maximization, and robustness optimization problems in the networks.
% Numerical experiments conducted on synthetic and real-world networks to evaluate our method, the efficacy of which is validated in epidemic prevalence and threshold prediction tasks. In addition, we discuss the potential impact of our method in immunization, influence maximization and robustness optimization problem in the networks.

%% file: texs/sec_intro.tex
\section{Introduction}
Predicting epidemic processes and determining epidemic thresholds stand at the core of epidemic research on networks. On the one side, forecasting the epidemic status—whether at the node-wise or the system-wide scale—plays a crucial role in many network optimization problems \cite{pastor2002immunization,IMP_2009,IMP_2018,Nonbacktracking_2021}, such as  influence maximization and network immunization. The epidemic threshold indicates the direction of the epidemic processes. Properly pinpointing this threshold is paramount for developing efficient network control strategies and interventions. For instance, in the context of epidemic spreading, an accurate prediction of the epidemic threshold is important in formulating preemptive vaccination strategies. 

% Although Monte Carlo simulations (MC) method can numerically forecast epidemic processes, it struggles to scale up to larger networks and fail to offer insights into the interplay between dynamics and network structure., analytical theoretical methods do.
Although one can utilize Monte Carlo (MC) simulations to numerically forecast epidemic processes, MC struggles to scale up to large networks and fails to offer theoretical insights. 
Thus, we require analytical theoretical methods with acceptable computational complexity to explore the interplay between epidemic dynamics and network structure.
The complexity of network structures and the stochastic nature of epidemic dynamics make the exact theoretical analyses of real epidemic networks intrinsically infeasible \cite{IMP_2009,Fundamental_2022}. 
Therefore, over the past two decades, many theoretical approximation methods \cite{Unification_Wang_2017,Fundamentals-Exact} have been developed to bridge the gap, including the quenched mean-ﬁeld theory (QMF) \cite{1238052,10.1145/1284680.1284681,4549746}, the pair quenched mean-ﬁeld (PQMF) theory \cite{Mata_2013,PhysRevResearch.1.033024,PhysRevE.102.012313} and message-passing (MP) based theory. Among various approximation methods, the Dynamic Message Passing (DMP) method is a prominent theoretical framework for epidemic dynamics analyzing \cite{PhysRevE.82.016101, DMP-SIR, newman2023message}, which was first introduced in \cite{PhysRevE.82.016101} to approximate the SIR model by utilizing the unidirectional dynamics of the model. 
Subsequent studies have extended DMP to approximate other unidirectional propagation dynamics \cite{PhysRevE.89.022805,DMP-SIR}. 
However, when dealing with recurrent epidemics, like the SIS and SIRS models, extension is not straightforward. 
By integrating the \textit{dynamic message} concept from DMP  with the mean-field method, Shrestha et al. \cite{Shrestha_rDMP} proposed a modified version dubbed the recurrent Dynamic Message Passing (rDMP) method for recurrent epidemics. 
A crucial feature of rDMP is its ability to prevent the \textit{echo chamber effect} observed along edges. This necessity has been empirically validated through experiments on both synthetic and real-world networks \cite{Shrestha_rDMP,Relevance_PRE_2018, Comparison_PRE_2022}.
Furthermore, by incorporating the non-backtracking (NB) matrix \cite{HASHIMOTO1989211,NB_PNAS_2013} to characterize the network structure, rDMP offers a new estimation of the epidemic threshold.

Despite the fact that the original rDMP theory is intuitive, it remains somewhat ad-hoc, leading to a probabilistic inconsistency flaw \cite{Shrestha_rDMP, PhysRevE.105.024308}. Therefore, there is a urgent need for a principled derivation, not only to address this flaw but also to deepen our understanding of the rDMP theory.
Furthermore, same as other message-passing (MP) methods, rDMP operates on the assumption that networks possess a locally tree-like structure \cite{newman2018networks}. However, real-world networks is clustering \cite{WS_clusterig, newman2003structure}, represented by existence of short loops like triangles.
Recent advances in MP research have highlighted notable performance improvements when these local structures are considered into the message-passing process \cite{GBP_2000, PhysRevE.96.042303, MPwL_2019, BPwL_SA_2021, gao2022neural, newman2023message}. 
Such results hint at the potential benefits rDMP could gain by adopting a similar methodology.
Nevertheless, Castellano et al. \cite{Relevance_PRE_2018} emphasize the critical role of the missing backtracking mechanism within the rDMP theory, particularly in the context of SIS dynamics in heterogeneous networks \cite{PhysRevE.94.042308, istvan2019mathematics, pastor2001epidemic}. When the degree exponent is large and the dynamics are predominantly steered by self-sustained activations from hub nodes \cite{PhysRevLett.111.068701}, the performance of rDMP is diminished. This result underscores the subtle balance rDMP must confront between preventing the echo chamber effects and retaining self-sustained activations.

In this work, to gain a comprehensive understanding of the rDMP theory, we systematically derived the dynamic message update equations, correcting the inherent flaw in the original rDMP. Our results align with the results based on cavity master equations (CME) \cite{PhysRevE.105.024308}. 
Interestingly, we observe the echo chamber effect not only on edges but also within local loops. This phenomenon results in the corrected rDMP still overestimating the probabilities of infections. By employing the same derivation, we find that the rDMP framework can be augmented to include these local loops.
Specifically, we introduce higher-order dynamic messages to counteract the echo chamber effect present in these loops.
What's more, when focusing exclusively on short loops, i.e., triangles, our refined rDMP theory offers a new theoretical estimation of the epidemic threshold.
Numerical experiments on synthetic and real-world networks validate the effectiveness of our theory in predicting epidemic prevalence and threshold. Our results highlight the benefits of limiting backtracking propagation not only along the network edges, but also along the local loops.

The rest of this paper is organized as follows. Section 2 introduces recurrent epidemic dynamics and existing theoretical approximation methods. Section 3 describes the echo chamber effect in local loops. Sections 4 and 5 rigorously derive our method and compare it with existing methods. Section 6 reports numerical experiments for predicting prevalence and threshold. Section 7 discusses and summarizes the entire paper.

%% file: texs/sec_othermethods.tex
\section{Recurrent Epidemics on Networks}

Consider an undirected contagion network, denoted as $\mathcal{G}=(\mathcal{N},\mathcal{E})$, with nodes $\mathcal{N}=\{1,2,\dots, N\}$ and edges $\mathcal{E}=\{(i,j)| i,j\in \mathcal{N} \text{ and } i\neq j \}$, and $|\mathcal{E}|=E$, the conventional stochastic compartmental models describe the epidemic states of nodes at time $t$ using discrete variables $\mathbf{x}^t=\{x_1^t,\dots,x_N^t\}$, of which the evolution is governed by a Markovian stochastic dynamics. Using the SIS model as a representative example, each state $x_i^t$ can either be susceptible ($S$) or infectious ($I$) for every $i\in\mathcal{N}$. Disease transmission through any edge occurs at a rate $\beta \geq 0$, while infectious nodes recover to a susceptible state at a rate $\gamma\geq 0$.

One aspect of epidemic dynamics is their directionality, which is crucial for the developing of their analytical method. Unidirectional models, such as SI and SIR, allow for the dynamical trajectory of a node to be concisely represented by a small set of variables. This efficient representation sidesteps the challenges posed by exponential complexity \cite{PhysRevE.82.016101,DMP-SIR} of the configuration space, laying the foundation for their effective DMP method derived from the cavity method. However, this form of simplification is not feasible for recurrent-state models like SIS and SIRS, wherein nodes can continually transition between states  $S\to I \to S$ (or $S\to I \to R \to S$). 

In this study, we focus our attention on recurrent-state dynamics, using the SIS model as a representative example model. Notably, our results can be extend to other recurrent-state models.

\paragraph{Exact Equations}
To enhance clarity, let $P(X_i^t) = P(x_i^t = X)$, where $X \in \{S, I\}$. The \textit{exact} temporal evolution of the marginal probability for node $i \in \mathcal{N}$ is \cite{Fundamentals-Exact}:
\begin{equation}\label{equ:sis_exact}
    \odv{P(I_i^t)}{t} = -\gamma P(I_i^t) + \beta \sum_{j \in \partial_i} P(S_i^t I_j^t),
\end{equation}
with the constraint that $P(S_i^t) + P(I_i^t) = 1$. Here, $\partial_i = \{j | (i, j) \in \mathcal{E}\}$ represents the set of neighbors for node $i$, and $P(S_i^t I_j^t)$ is the joint probability of states $S_i^t$ and $I_j^t$.

While Eq.(\ref{equ:sis_exact}) is exact, it isn't closed, owing to its dependence on the solution for $P(S_i^t I_i^t)$. Notably, the exact evolution of $P(S_i^t I_j^t)$ dependents on the joint probabilities involving three node states, introducing a hierarchy of dependencies. This ends in a system characterized by $2^N$ equations, making it computationally infeasible for real-world networks. To address this complexity and achieve a closed system, several theoretical approximation strategies have been developed, delicately balancing between computational overhead and analytical precision \cite{Unification_Wang_2017,Fundamentals-Exact}.

\paragraph{Approximation Methods}
Among the various approximation techniques, the quenched mean-field (QMF) is the most popular one due to its simplicity \cite{QMF_PhysRevE.89.052802,QMF_2009,QMF_MMCA_2010}. QMF assumes independence between all pairs of nodes, approximating $P(S_i^t I_j^t)$ as $P(S_i^t)P(I_j^t)$. This results in a closed system of equations, expressed as:
\begin{equation}\label{equ:sis_qmf}
    \odv{P(I_i^t)}{t} =  -\gamma P(I_i^t) + \beta P(S_i^t)\sum_{j\in\partial_i} P(I_j^t).
\end{equation}

An advanced method involves accounting for the dynamic correlations between node pairs, which QMF neglects. By employing the pair-approximation (PA) \cite{PhysRevE.85.056111, Mata_2013}, the three-point probability can be approximated by two-point probabilities:
\begin{equation}\label{equ:pa_pa}
    P(x_i^t x_j^t x_k^t) \approx \frac{P(x_i^t x_j^t)P(x_j^t x_k^t)}{P(x_j^t)}.
\end{equation}
Using the PA technique (sometimes referred to as moment closure), one can capture the dynamic correlations between node pairs, which is represented in the evolution of the two-point probability as:
\begin{equation}\label{equ:pa_2}
    \odv{P(S_i^t I_j^t)}{t} = -2\gamma P(S_i^t I_j^t) + \gamma P(I_j^t)+\beta\sum_{k\in\partial_j\backslash i}\frac{P(S_i^t S_j^t)P(S_j^t I_k^t)}{P(S_j^t)} -\beta\sum_{k\in\partial_j\backslash i}\frac{P(S_i^t I_j^t)P(I_j^t I_k^t)}{P(I_j^t)},
\end{equation}
where $\partial_j\backslash i$ represents the set of neighbors of node $j$ excluding node $i$, and $\sum_{X,Y\in{S,I}}P(X_i^tY_j^t) = 1$. The equations given by Eq.(\ref{equ:sis_exact}) and Eq.(\ref{equ:pa_2}) collectively form a closed system, under the assumption that the underlying network structure has a locally tree-like characteristic \cite{newman2018networks}. Notably, an equivalent perspective to PA, as shown in \cite{burgio_network_2021}, begins by breaking down the probability $P(S_i^t I_j^t)$ into $P(S_i^t)P(I_j^t\mid S_i^t)$, and the evolution of the resulting conditional probability is derived by Bayes' theorem and the Markov property of the dynamics.

However, both QMF and PA are vulnerable to the \textit{echo chamber effect in edges}, potentially overestimating the node infection probability. Drawing inspiration from the message-passing approach of the SIR model \cite{PhysRevE.82.016101}, Shrestha et al. \cite{Shrestha_rDMP} introduced dynamic messages into directed edges, effectively preventing the echo chamber effect. This led to a new approximation $P(S_i^t I_j^t) \approx P(S_i^t)P_{\to i}(I_j^t)$, and the resulting system, termed recurrent dynamic message-passing (rDMP), defined by:
\begin{align}
    \odv{P(I_i^t)}{t} &=  -\gamma P(I_i^t) + \beta P(S_i^t)\sum_{j\in\partial_i} P_{\to i}(I_{j}^t), \label{equ:rdmp_1}\\
    \odv{P_{\to i}(I_{j}^t)}{t} &=  -\gamma P_{\to i}(I_{j}^t) + \beta P(S_j^t)\sum_{k\in\partial_j \backslash i} P_{\to j}(I_k^t), \label{equ:rdmp_2}
\end{align}
where $P_{\to i}(I_j^t)$ represents the \textit{cavity} probability, which is defined on the directed edge $j$ to $i$, indicating that node $j$ is infected at time $t$ when node $i$ is absent. It is also noteworthy that $P_{\to i}(I_j^t) + P_{\to i}(S_j^t) = 1$.  $P_{\to i}(I_j^t)$ is referred to as a \textit{dynamic message} from $j$ to $i$. Nevertheless, the rDMP was derived in an intuitive, \textit{ad hoc} manner, leading to a probabilistic flaw, i.e., the contribution of term $P(S_j^t)$ in Eq.(\ref{equ:rdmp_2}) is not consistent with the evolution of the corresponding cavity probability $P_{\to i}(I_j^t)$. 
Recently, Ortega et al. \cite{PhysRevE.105.024308} rigorously derived the rDMP theory using cavity master equations (CME) and corrected this flaw (term marked by a box in Eq.(\ref{equ:cme})):
\begin{equation}\label{equ:cme}
    \odv{P_{\to i}(I_{j}^t)}{t} =  -\gamma P_{\to i}(I_{j}^t) + \beta \boxed{P_{\to i}(S_j^t)} \sum_{k\in\partial_j\backslash i} P_{\to j}(I_k^t).
\end{equation}
The effectiveness of this correction is validated through numerical experiments presented in \cite{PhysRevE.105.024308}. Notably, CME and rDMP are essentially the same method, they improve the basic mean-field theory by preventing the echo chamber effect (only) along edges.
In the subsequent sections, the same correction made in CME will be derived in a different way, based on which, we will propose an extended method of both rDMP and CME.

%% file: texs/sec_echochamber.tex
\section{Echo Chamber Effects in Local Loops}
As introduced in the previous section, either the rDMP or the CME\footnote{As the CME is a revision of the rDMP, we will use the term `rDMP' to refer to both methods unless otherwise stated.} only prevents the echo chamber effect in the edges, which is equivalent to a directed loop of length 2. In the following, we will demonstrate that the phenomenon of echo chamber effect is not confined to edge, but exits in local loops of greater lengths as well.

\begin{figure}[ht]
    \centering
    \includegraphics[width=0.8\textwidth]{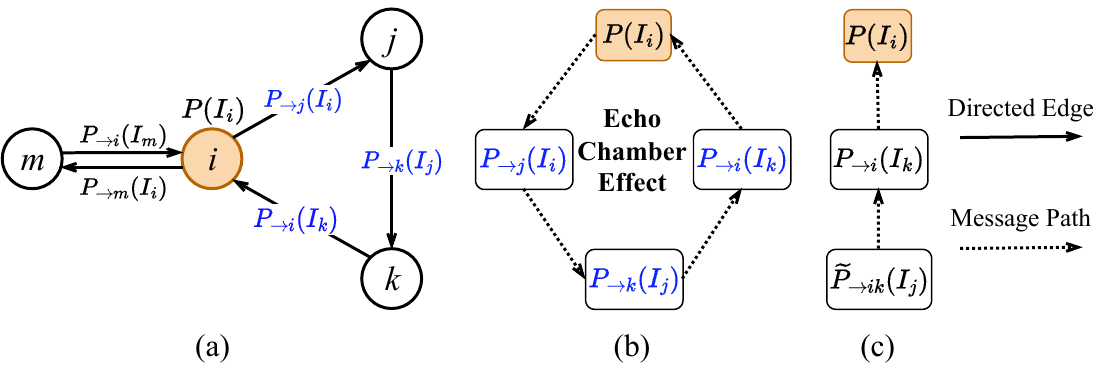}
    \caption[Echo Chamber Effect in Local Loops]{\textbf{Echo chamber effect in local loops.} (a) A simple case where node $i$ acts as the epidemic source, and there exits two loops: $(m,i)$ and $(i,j,k)$. (b) With the application of rDMP update equation, the local loop $(i,j,k)$ engenders an echo chamber effect to the marginal probability $P(I_i)$. (c) The introduction of a higher-order dynamic message breaks down the self-amplification cycle induced by the loop $(i,j,k)$.}
    \label{fig:echo_chamber_effect}
\end{figure}

Consider a simple case, as pictured in Fig. \ref{fig:echo_chamber_effect}(a), where node $i$ is initially infectious with probability $P(I_i^0)=0.5$ and others with $0$, and the transmission rate $\beta=1.0$ while recovery rate $\gamma=0$. Here, we focus on two local loops that node $i$ lies in, namely, $(i,m)$ of length 2 and $(i,j,k)$ of length 3. Clearly, by utilizing the messages $P_{\to i}(I_m)$ and $P_{\to m}(I_i)$, the rDMP effectively counters any unintended amplification of infectiousness between nodes $m$ and $i$ \cite{Shrestha_rDMP}. 

However, following the updating equation of rDMP, node $i$ undergoes self-amplification in the loop $(i,j,k)$, as shown in Fig. \ref{fig:echo_chamber_effect}(b). Specifically, in this setting, rDMP is simplified as 
$dP(I_{i}^t)/dt=P(S_i^t)\left[P_{\to i}(I_j^t)+P_{\to i} (I_k^t)\right],$ and the dynamic messages within are governed by $d P_{\to i}(I_{j}^t)/dt = P_{\to i}(S_{j}^t)\left[P_{\to j}(I_i^t)+P_{\to j} (I_k^t)\right],$ which eventually reaching positive values and causing $P(I_i^t)>P(I_i^0)$. Given that only node $i$ is the initial source of propagation, such a result violates the causal constraint that $\forall t>0, P(I_i^t)\leq P(I_i^0)$. 

Therefore, besides the edges addressed by rDMP, all other local loops can also lead to an overestimation of the marginal probabilities. A loop of length \(L\) induces self-amplification after \(L\) time steps, yet this effect diminishes exponentially as length \(L\) increases. This observation indicates that the echo chamber effect of relatively long loops is negligible, which explains why rDMP can still provide accurate approximations even without considering loops with \(L\geq 3\).  Guided by this, we specifically focus on addressing local loops of length 3 and intentionally overlook the effects of longer loops.

%% file: texs/sec_lrdmp.tex
\section{Message Passing with Local Loops}
Same as CME \cite{PhysRevE.105.024308}, we seek for a principled and rigorous derivation of rDMP theory. It is encouraging to observe in the following content that this derivation process not only corrects the flaw of rDMP theory but also organically introduces novel higher-order messages to prevent self-amplification within local loops caused by echo chamber effect.

The essence of the rDMP lies in utilizing the dynamic messages to approximate the two-point probability, i.e., $P(S_i^t I_j^t) \approx P(S_i^t)P_{\to i}(I_j^t)$. A key observation is that $P_{\to i}(I_j^t)$ is equivalent to the node-wise probability in the  cavity graph $\mathcal{G}_{\to i}$ (same as $\mathcal{G}$ except node $i$ is removed) . This equivalence implies that the evolution of $P_{\to i}(I_j^t)$ within $\mathcal{G}_{\to i}$ cab be derived in the same principle as in  Eq.(\ref{equ:rdmp_1}):
\begin{align}
\odv{P_{\to i}(I_{j}^t)}{t} 
& =  -\gamma P_{\to i}(I_{j}^t) + \beta\sum_{k\in\partial_j \backslash i} P_{\to i}(S_j^t I_k^t)\\
& \approx -\gamma P_{\to i}(I_{j}^t) + \beta P_{\to i}(S_j^t)\sum_{k\in\partial_j \backslash i} \tilde{P}_{\to ij}(I_k^t). \label{equ:lrdmp_1}
\end{align}    
where $P_{\to i}(S_j^t I_k^t) \approx P_{\to i}(S_j^t)\tilde{P}_{\to ij}(I_k^t)$. The newly introduced \textit{higher-order}\footnote{Here, `higher-order' refers to the dynamic message or cavity probability as more nodes are omitted from consideration.} message $\tilde{P}_{\to ij}(I_k^t)$ denotes the cavity probability of $I_k^t$ within $\mathcal{G}$, with  nodes $i$ and $j$ removed. It is also equivalent to the cavity probability of $I_k^t$ within $\mathcal{G}_{\to i}$ in the absence of node $j$. One significant feature of this update equation, as indicated by Eq.(\ref{equ:lrdmp_1}), is its capacity to explicitly bypass the echo chamber effect in loops, as pictured in Fig. \ref{fig:echo_chamber_effect}(c).

Bearing in mind that we are ignoring all propagation paths with lengths more than 3. In the case where $k\notin \partial_i$, the self-amplification paths of node $i$ through $k$ must have lengths of at least 4, which implies that $\tilde{P}_{\to ij}(I_k^t)$ can be reduced to lower-order message $P_{\to j}(I_k^t)$. Otherwise, $\tilde{P}_{\to ij}(I_k^t)$ must be retained to prevent propagation in the loop $(i,j,k)$. Formally, given $j\in \partial_i$ and $k \in \partial_j\backslash i$, message $\tilde{P}_{\to ij}(I_k^t)$ can be reduced according to the local structure as:
\begin{equation}\label{equ:lrdmp_reduce1}
    \tilde{P}_{\to ij}(I_k^t) = 
    \begin{cases}
    P_{\to j}(I_k^t) & \text{if } k \notin \partial_i, \\
    P_{\to ij}(I_k^t)   & \text{otherwise}.
  \end{cases}
\end{equation}

To obtain a closed evolution system, we once again interpret message $P_{\to ij}(I_k^t)$ as the node-wise probability within the cavity graph $\mathcal{G}_{\to ij}$, and thereby we arrive at the following:
\begin{align}
\odv{P_{\to ij}(I_k^t)}{t} 
& = -\gamma P_{\to ij}(I_k^t) + \beta \sum_{l\in\partial_{k}\backslash ij} P_{\to ij}(S_k^tI_l^t) \\
& \approx  -\gamma P_{\to ij}(I_k^t) + \beta P_{\to ij}(S_k^t) \sum_{l\in\partial_{k}\backslash ij} \tilde{P}_{\to ijk}(I_l^t),
\end{align}
and the message $\tilde{P}_{\to ijk}(I_l^t)$ denotes the probability of $I_l^t$ in the absence of nodes $i$,$j$ and $k$. Note that the necessity of $\tilde{P}_{\to ijk}(I_l^t)$ implies that nodes $i,j,k$ form a triangle and $l\in\partial_k$. Utilizing the same manner used for the reduction of $\tilde{P}_{\to ij}(I_k^t)$, the higher-order message $\tilde{P}_{\to ijk}(I_l^t)$ can be reduced to lower-order ones as:
\begin{equation}\label{equ:lrdmp_reduce2}
    \tilde{P}_{\to ijk}(I_l^t) = 
    \begin{cases}
        P_{\to k}(I_l^t) & \text{if } l\notin \partial_j\text{ and } l\notin \partial_i\\
        P_{\to ik}(I_l^t) & \text{if } l\notin \partial_j\text{ and } l\in \partial_i\\
        P_{\to jk}(I_l^t) & \text{if } l\in \partial_j\text{ and } l\notin \partial_i\\
        P_{\to ijk}(I_l^t) & \text{if } l\in \partial_j\text{ and } l\in \partial_i.
    \end{cases}
\end{equation}
In the case where $l\in \partial_j\text{ and } l\in \partial_i$, the $P_{\to ijk}(I_l^t)$ can not be reduced, therefore, we adopt a heuristic  pair-approximation to close the system as follows:
\begin{equation}\label{equ:lrdmp_pa}
    P_{\to i j k}\left(I_l^t\right) \approx \min \left( \left\{P_{\to i j}(I_l^t), P_{\to j k}(I_l^t), P_{\to i k}(I_l^t) \right\} \right).
\end{equation}
The intuition behind Eq.(\ref{equ:lrdmp_pa}) is the observation that $P_{\to i j k}\left(I_l^t\right)$ is upper bounded by those lower-order messages. It is worth emphasizing that Eq. (\ref{equ:lrdmp_pa}) is not the only choice for approximating $P_{\to i j k}\left(I_l^t\right)$ using lower-order messages, but it possesses permutation invariance w.r.t the order of nodes. 

At this point, in principle, a closed system considering local loops (with length $L<4$) is derived, termed as \textbf{L}oop corrected \textbf{r}ecurrent \textbf{D}ynamic \textbf{M}essage \textbf{P}assing approach, or \textbf{LrDMP}. Integratedly, the LrDMP equations for the SIS model take the following form:
\begin{equation}\label{equ:lrdmp}
  \begin{aligned}
    \odv{P(I_i^t)}{t} & =  -\gamma P(I_i^t) + \beta P(S_i^t)\sum_{j\in\partial_i} P_{\to i}(I_{j}^t) \\
    \odv{P_{\to i}(I_{j}^t)}{t} & = -\gamma P_{\to i}(I_{j}^t) + \beta P_{\to i}(S_j^t)\sum_{k\in\partial_j \backslash i} \tilde{P}_{\to ij}(I_k^t) \\
    \odv{P_{\to ij}(I_k^t)}{t} & = -\gamma P_{\to ij}(I_k^t) + \beta P_{\to ij}(S_k^t) \sum_{l\in\partial_{k}\backslash ij} \tilde{P}_{\to ijk}(I_l^t)
  \end{aligned}
\end{equation}
where the dynamic messages $\tilde{P}_{\to ij}(I_k^t)$ and $\tilde{P}_{\to ijk}(I_l^t)$ follow the structural reduction and heuristic pair-approximation respectively in Eqs. (\ref{equ:lrdmp_reduce1}), (\ref{equ:lrdmp_reduce2}) and (\ref{equ:lrdmp_pa}). Dynamic messages $\{P_{\to (\cdot)}(I^0_i)\}_{i\in\mathcal{N}}$ can be initialized by the corresponding initial probabilities as $P_{\to (\cdot)}(I^0_i)=P(I^0_i)$. The total number of equations in system Eq.(\ref{equ:lrdmp}) is $N+2E+3T$,where $T$ is the number of all triangles in $\mathcal{G}$.

Generalizing LrDMP to other recurrent-state models, such as SIRS and SEIS, can be achieved with ease. Additionally, this system can be readily extended to address the case of heterogeneous recovery and transition rates. Even for time-dependent or neighborhood state-dependent epidemic models, these equations are applicable with certain modifications \cite{Shrestha_rDMP}. While it might be superfluous, as previously noted, LrDMP can indeed be expanded to accommodate loops that exceed a length of 3 in a consistent manner.

To our knowledge, LrDMP is the first effort to incorporate local loops within the dynamic message-passing paradigm, offering a more delicate description of contagion network structures.

\paragraph{Relation to Existing Methods} 
Clearly, when graph $\mathcal{G}$ contains no local loops of length 3 (i.e., triangle),  the equation $\tilde{P}_{\to ij}(I_k^t)=P_{\to j}(I_k^t)$ always holds true. 
In this circumstance, Eq.(\ref{equ:lrdmp_1}) naturally reduces to Eq.(\ref{equ:cme}) in CME. 
Such an observation implies that CME is a special and simplified version of the proposed LrDMP.
What's more, this also  means rDMP corrects the flaw in original rDMP in a principled way.
On the other hand, a potential limitation of CME appears: its persistent reliance on the pair approximation $P_{\to ij}(I_k^t)\approx P_{\to j}(I_k^t)$, an approximation that might be oversimplified for abundant real-world networks with many local loops. 

As stated above, CME is a variant of LrDMP only addressing shorter local loops (edge). Therefore, to explicitly probe the influence of loop length addressed, it would be convenient to refer to these methods as \textbf{LrDMP-1} and \textbf{LrDMP-2}, respectively. This naming convention will be adopted in the numerical experiments presented later in the paper.

\paragraph{Limitations}
It is crucial to restate that rDMP–like methods have inherent limitations, especially for SIS models as argued in \cite{Relevance_PRE_2018, PhysRevE.105.024308}. rDMP prevents the echo chamber effect  by using the non-backtracking mechanism, i.e., nodes can not propagate the disease back immediately. 
However, in sparse graphs, especially scale-free ones, the backtracking transition serves as the foundation for the activation of the epidemic state: hub nodes maintain the network epidemic state by continuously getting infected by their neighbors and subsequently reinfecting them in return.
Thus, rDMP-like methods will certainly fail in tree graphs and scale-free graphs. 
Conversely, as previously noted, echo chamber propagation can cause theoretical methods to overestimate infection states. 
Therefore, understanding how non-backtracking mechanism impact the approximation efficacy of message-passing methods in real-world networks is still an open question.

\section{Linear Stability Analysis and Epidemic Threshold}
The nonlinearity of the proposed LrDMP system poses challenges for deriving an analytical solution. Consequently, it is common practice to probe the epidemic threshold of the system using the standard linear stability analysis. 

Clearly, as described by Eq.(\ref{equ:lrdmp}), a naive absorbing state of the system is attained when all marginals and dynamic messages are zero. At this point, the system is deemed healthy with the absence of any disease. To perturb around this equilibrium point, by assigning minimal values \(0 < \epsilon_{i\to (\cdot)} \ll 1\) to each message \(P_{\to (\cdot)}(I_i) = 0\), the LrDMP system can be linearized in the subsequent manner:
\begin{equation}\label{equ:lrdmp_linear1}
    \begin{aligned}
        \odv{\epsilon_{j\to i}}{t} & = -\gamma \epsilon_{j\to i} + \beta \sum_{k\in\partial_j \backslash i}  \left(\epsilon_{k\to j}\mathbbm{1}_{k\notin\partial_i} + \epsilon_{k\to ij}\mathbbm{1}_{k\in\partial_i}\right)
    \end{aligned}
\end{equation}
\begin{equation}\label{equ:lrdmp_linear2}
    \begin{aligned}
        \odv{\epsilon_{k\to ij}}{t}  =  -\gamma \epsilon_{k\to ij} + \beta \sum_{l\in\partial_{k}\backslash ij} & \left[ \epsilon_{l\to k}\mathbbm{1}_{l\notin\partial_i}\mathbbm{1}_{l\notin\partial_j} + \epsilon_{l\to ik}\mathbbm{1}_{l\in\partial_i}\mathbbm{1}_{l\notin\partial_j}+\right.  \\
        & \left. \epsilon_{l\to jk}\mathbbm{1}_{l\notin\partial_i}\mathbbm{1}_{l\in\partial_j} + \underline{\epsilon_{l\to rs}}\mathbbm{1}_{l\in\partial_i}\mathbbm{1}_{l\in\partial_j}\right] 
    \end{aligned}
\end{equation}
where $\mathbbm{1}$ is the indicator function. Note that the pair-approximation used in Eq.(\ref{equ:lrdmp_pa}) renders linearization inherently infeasible, thus we employ a randomization trick to circumvent this obstacle. Specifically, when $l\in\partial_i $ and $l\in\partial_j$, we adopt $\epsilon_{l\to ijk}\approx\epsilon_{l\to rs}$ where $r,s$ is randomly chosen from $\{i,j,k\}$ without replacement, as shown in the  last term of Eq.(\ref{equ:lrdmp_linear2}). This new approximation is used only to derive the linearization system, but not in the prevalence prediction. As shown in the subsequent numerical experiments on real-world networks, the influence of the newly introduced random operation on threshold prediction is neglectable.

\begin{figure}
    \centering
    \includegraphics[width=0.6\textwidth]{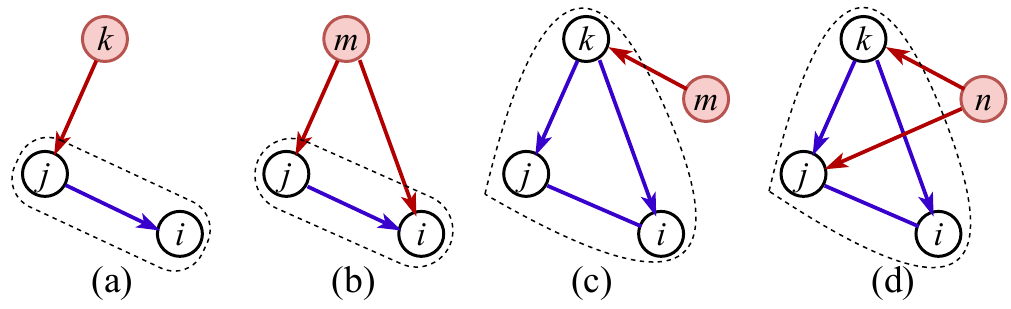}
    \caption{\textbf{Non-backtracking paths between directed edges and triangles.} 
    (a) Edge to edge (matrix $\mathbf{B}$): $(k\to j)\to (j\to i)$.
    (b) Triangle to edge (matrix $\mathbf{C}$): $(m\to ij)\to (j\to i)$.
    (c) Edge to triangle (matrix $\mathbf{D}$): $(m\to k)\to (k\to ij)$.
    (d) Triangle to triangle (matrix $\mathbf{E}$): $(n\to jk)\to (k\to ij)$.}
    \label{fig:nb_relations}
\end{figure}

Let $\epsilon_{edge}=[\cdots,\epsilon_{j\to i},\cdots]\in \mathbb{R}^{2E}$ represent messages on all directed edges, and $\epsilon_{tri} = [\cdots,\epsilon_{k\to ji},\cdots]\in \mathbb{R}^{3T}$ represent messages on all \textit{directed} \footnote{We define the directionality of a triangle by selecting one node as the source and the other two together as the target.} triangles. Then, a matrix form of the system in Eqs.(\ref{equ:lrdmp_linear1}) and (\ref{equ:lrdmp_linear2}) is:
\begin{equation}\label{equ:lrdmp_linear}
    \odv{\Vec{\epsilon}}{t} = 
    \left(\beta\begin{bmatrix}
        \mathbf{B} & \mathbf{C} \\
        \mathbf{D} & \mathbf{E} \\
    \end{bmatrix}-\gamma\mathbf{I}\right)\Vec{\epsilon} % = \left(\beta\mathbf{T} -\gamma\mathbf{I}\right)\Vec{\epsilon}
\end{equation}
where $\Vec{\epsilon}=[\epsilon_{edge},\epsilon_{tri}]^T$. The submatrixes $\mathbf{B}\in\mathbb{R}^{2E\times 2E},\mathbf{C}\in\mathbb{R}^{2E\times 3T},\mathbf{D}\in\mathbb{R}^{3T\times 2E}$ and $\mathbf{E}\in\mathbb{R}^{3T\times 3T}$ indicate the \textit{non-backtracking} paths between edges and triangles, defined as follows:
\begin{equation}
\begin{aligned}
\mathbf{B}_{j \rightarrow i, l \rightarrow k} & =\delta_{k, j}\left(1-\delta_{l, i}\right) \mathbbm{1}_{l \notin \partial_i} \\
\mathbf{C}_{j \rightarrow i, m \rightarrow kl} & =\delta_{\{i, j\},\{k, l\}} \\
\mathbf{D}_{k \rightarrow ij, m \rightarrow l} & =\delta_{l, k}\left(1-\delta_{m, i}\right)\left(1-\delta_{m, j}\right) \mathbbm{1}_{m \notin \partial_i} \mathbbm{1}_{m \notin \partial_j} \\
\mathbf{E}_{k \rightarrow ij, n \rightarrow lm} & =\mathbbm{1}_{n\notin\{i,j,k\}}\mathbbm{1}_{\{l,m\}\subset\{i,j,k\}}\left(\mathbbm{1}_{n \notin \partial_j} + \mathbbm{1}_{n \notin \partial_i} +\mathbbm{1}_{n \in \partial_i} \mathbbm{1}_{n \in \partial_j} \mathcal{R}\right)
\end{aligned}
\end{equation}
where $\mathcal{R} \in \{0,1\}$, represents whether $\epsilon_{n\to lm}$ is the selected message in the random choice step. These submatrices encompass more than just the edges; they capture the non-backtracking paths among all local loops with lengths shorter than four, i.e., edges and triangles. These relationships are visualized in Fig. \ref{fig:nb_relations}. It is crucial to note that the submatrix $\mathbf{B}$ is distinct from the well known non-backtracking ($\mathbf{NB}$) matrix\cite{HASHIMOTO1989211}, as $\mathbf{B}$ exclusively considers non-backtracking relations for pairs of edges that do not belong to a common triangle.  Collectively, these submatrices extend the non-backtracking relationship originally defined solely for edges, and we refer to the resulting new matrix as the \textit{triangular non-backtracking matrix} ($\mathbf{TNB}$):
\begin{equation}
    \mathbf{TNB} = \begin{bmatrix}
        \mathbf{B} & \mathbf{C} \\
        \mathbf{D} & \mathbf{E} \\
    \end{bmatrix}
\end{equation}

A similar matrix was proposed by Zhang \cite{PhysRevE.96.042303} in the study of message-passing methods for percolation problems. However, there is a significant difference between them: while Zhang's matrix only considers triangles that do not share any edge, our method encompasses all triangles.

Since every element of matrix $\mathbf{TNB}$ is non-negative, Perron-Frobenius theorem applies and we can conclude that its largest eigenvalue $\Lambda_{\mathbf{TNB}}\geq 0$. A new estimate of the epidemic threshold $\lambda_c$ is then derived by the standard linear stability analysis of Eq.(\ref{equ:lrdmp_linear}):
\begin{equation}
    \lambda_c = \frac{\beta}{\gamma} = \Lambda_{TNB}^{-1}.
\end{equation}
Unlike existing theoretical approaches that characterize network structure using the adjacency matrix or non-backtracking matrix, the LrDMP proposed here leverages the $\mathbf{TNB}$ matrix. This captures the local loops (e.g., triangles) in the network with greater detail, an aspect not considered by existing methods to the best of our knowledge.

%% file: texs/sec_experiments.tex
\section{Numerical Experiments}
In this section, we conduct numerical experiments to assess the effectiveness of the proposed approximation method, LrDMP-1 (i.e., CME) and LrDMP-2, for both epidemic prevalence prediction and threshold prediction tasks.

\subsection{Epidemic Prevalence}
Let the time-evolving epidemic prevalence be denoted by
 \begin{equation}
     \rho(t) = \frac{1}{N} \sum_{i=1}^{N} P(I_i^t),
 \end{equation}
 where its steady value, $\rho$, functions as the \textit{order parameter}  of the system. We evaluate the effectiveness of the proposed method LrDMP and compare it with other baseline approximation methods such as QMF, rDMP, and PA. The comparison is based on their prediction of epidemic prevalence against the results obtained from average of Monte Carlo (MC) simulations, which are regarded as the ground truth prevalence. The accuracy of predictions for each method is quantified using the $L_1$ error, defined as:
\begin{equation}
    L_1 = \frac{1}{N}\sum_{i}|P(I_i^t) - \hat{P}(I_i^t)|,
\end{equation}
where $P(I_i^t)$ represents the true prevalence at time $t$ for node $i$, and $\hat{P}(I_i^t)$ is the predicted prevalence. When computing the MC averages, simulations where the epidemic is wiped out during the early iterations were omitted. This exclusion is due to the inability of mean-field methods to address these fluctuations near the epidemic threshold. It is noticeable that in a finite system, $\lim_{t\to\infty}\rho(t) = 0$, therefore, we confine the evolution of the system to a predetermined number of finite time steps.

\begin{figure}[t]
    \centering
    \begin{subfigure}{0.48\textwidth}
        \includegraphics[width=\linewidth]{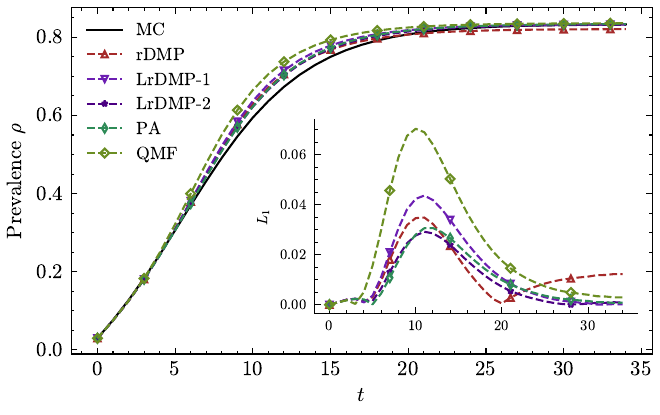}
        % \caption{First Image}
    \end{subfigure}
    \hfill
    \begin{subfigure}{0.48\textwidth}
        \includegraphics[width=\linewidth]{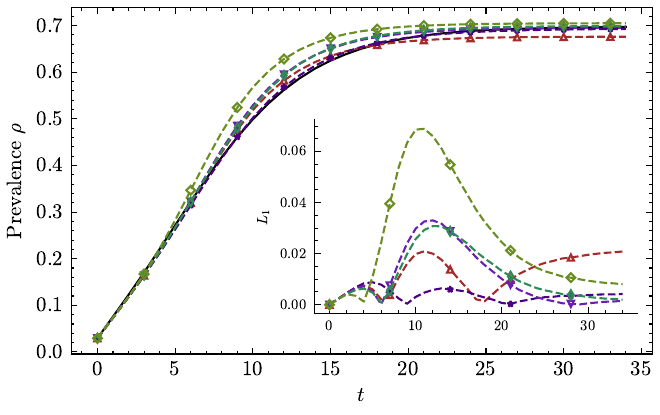}
        % \caption{Second Image}
    \end{subfigure}
    \caption{\textbf{Prevalence Prediction on the Karate Club Network.} Following the settings from \cite{Shrestha_rDMP}, node 0 serves as the epidemic origin. In the (\textbf{\textit{left}}) sub-figure, the parameters are set to $\beta=0.1$ and $\gamma=0.05$. Meanwhile, in the (\textbf{\textit{right}}) sub-figure, $\gamma$ is adjusted to $0.1$. The MC simulations represent the average of $10^5$ independent trials, and are compared against the predictions from approximate methods LrDMP-1, LrDMP-2, QMF, rDMP, and PA. The inset figures highlight the $L_1$ error between the MC results and the approximate methods.}
    \label{fig:prevalence_prediction_karate}
\end{figure}
\begin{figure}[t]
    \centering
    \begin{subfigure}{0.49\textwidth}
        \includegraphics[width=\linewidth]{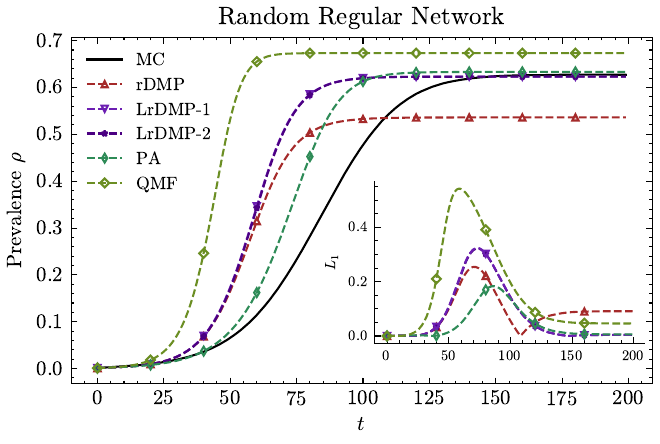}
        % \caption{First Image}
    \end{subfigure}
    \hfill
    \begin{subfigure}{0.49\textwidth}
        \includegraphics[width=\linewidth]{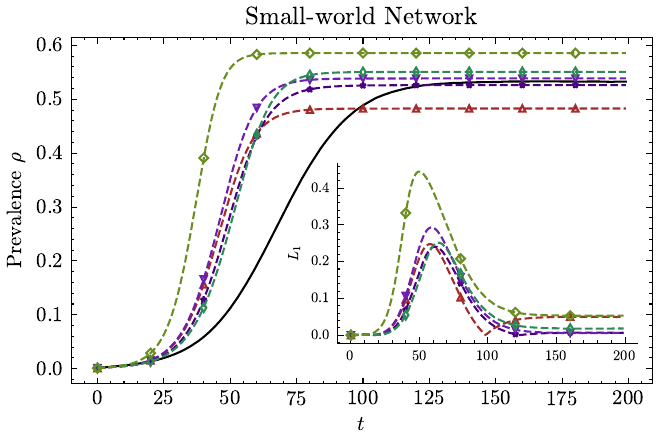}
        % \caption{Second Image}
    \end{subfigure}
    \caption{\textbf{Prevalence Prediction on Synthetic Network.} \textbf{(\textit{left})}: Results on a random regular network with $N=1000,k=3$. The epidemic outbreak starts from one node, with the parameters set as $\beta=0.11$ and $\gamma=0.1$, same as the settings in \cite{PhysRevE.105.024308}. \textbf{(\textit{right})}: For a small-world network generated by the Watts-Strogatz (WS) model \cite{WS_clusterig} with configuration $(N=1000, k=4, p=0.2)$, the epidemic begins from a single node, with parameters $\beta=0.1$ and $\gamma=0.15$. In both settings, the MC results are the average of $10^4$ independent simulations, juxtaposed against the predictions from approximate methods LrDMP-1, LrDMP-2, QMF, rDMP, and PA. The inset figures display the $L_1$ error between the MC simulations and the approximate methods.}
    \label{fig:prevalence_prediction_rrn_ws}
\end{figure}

\paragraph{Results}
Following the parameter settings from \cite{Shrestha_rDMP}, we first conduct our prevalence prediction experiments on the Karate network, as depicted in Fig. \ref{fig:prevalence_prediction_karate}. We use same transition rate of $\beta=0.1$ and explore two different recovery rates: $\gamma=0.05$ and $\gamma=0.1$. For both setups, the epidemic starts from node 0, evolving up to a maximum of 35 time steps. The MC is the average of $10^5$ independent simulations, excluding absorbed trials. The results in Fig. \ref{fig:prevalence_prediction_karate} clearly demonstrate that, irrespective of the dynamic parameter settings and across all time steps, LrDMP-1 and LrDMP-2 either outperform or are on par with other methods in terms of prevalence predictions. LrDMP-1 shows improved performance over rDMP by introducing corrections systematically. The improvement of LrDMP-2 over LrDMP-1 further highlights that considering the echo chamber effect along local loops in dense networks can refine prediction accuracy.

In Fig. \ref{fig:prevalence_prediction_rrn_ws}, we present simulations of the SIS model on two networks: a random regular network (RRN) and a small-world network (SWN). For the RRN, we use parameters $N=1000$ and average degree $k=3$, and for the SWN, $N=1000$, $k=4$, and a rewiring probability $p=0.2$. Both simulations begin with a single infectious node. The respective epidemic parameters are $\beta=0.11, \gamma=0.1$ for the RRN and $\beta=0.1, \gamma=0.15$ for the SWN. The RRN, characterized by its sparse and locally tree-like structure, lacks local loops such as triangles. Consequently, in this network, LrDMP-2 simplifies to LrDMP-1. This accounts for the identical prevalence predictions of LrDMP-1 and LrDMP-2 on the RRN. Conversely, the SWN displays significant local clustering and numerous triangles. As LrDMP-2 is more adept at addressing these local loops, it outperforms LrDMP-1 on the SWN. Consistent with results from the Karate network, the LrDMP methods surpass all baseline methods, including rDMP, across all time steps—particularly when determining the steady-state size of the epidemic.

\subsection{Epidemic Threshold}
The capability of a theoretical approach to precisely predict the epidemic threshold is an evidence to its effectiveness in capturing the essence of epidemic dynamics \cite{PhysRevE.86.041125, wang_predicting_2016, Comparison_PRE_2022}. In this section, we examine three theoretical predictions of epidemic thresholds: $\Lambda_{A}^{-1}$, $\Lambda_{NB}^{-1}$, and $\Lambda_{TNB}^{-1}$, where $\Lambda$ denotes the largest eigenvalue of a given matrix. Specifically, $\mathbf{A}$ represents the adjacency matrix associated with QMF theory. $\mathbf{NB}$ is the non-backtracking matrix associated with the rDMP and CME theory, and $\mathbf{TNB}$ is a novel matrix introduced by our LrDMP theory. 

\paragraph{Stochastic Simulation}
To obtain the numerical thresholds and then validate the efficacy of the theoretical predictions, we carry out stochastic simulations of the SIS model employing the Gillespie method \cite{Gillespie_COTA2017303}. These simulations consisted of $10^6$ steps for both relaxation and averaging time. To minimize memory overhead, the optimized quasistationary method \cite{QS_COSTA2021108046} is adopted to handle the absorbing states. 

For simplicity and without loss of generality, we set $\gamma=1$ for all simulations, which implies the effective infection rate, denoted as $\lambda$, is given by $\lambda=\beta/\gamma=\beta$. Let $\rho_{\lambda}$ represent the average epidemic prevalence computed in the quasi-stationary regime with parameter $\lambda$. The numerical epidemic threshold, $\lambda_c$, is determined by the infection rate $\lambda$ corresponding to the peak value of the dynamical susceptibility $\chi$, i.e., $\lambda_c = \arg\max_{\lambda}\chi$, where $\chi$ is defined as $\chi = N \left( \langle \rho_{\lambda}^2 \rangle - \langle \rho_{\lambda} \rangle^2 \right) / \langle \rho_{\lambda} \rangle$.

\begin{figure}[th]
    \centering
    \includegraphics[width=0.8\textwidth]{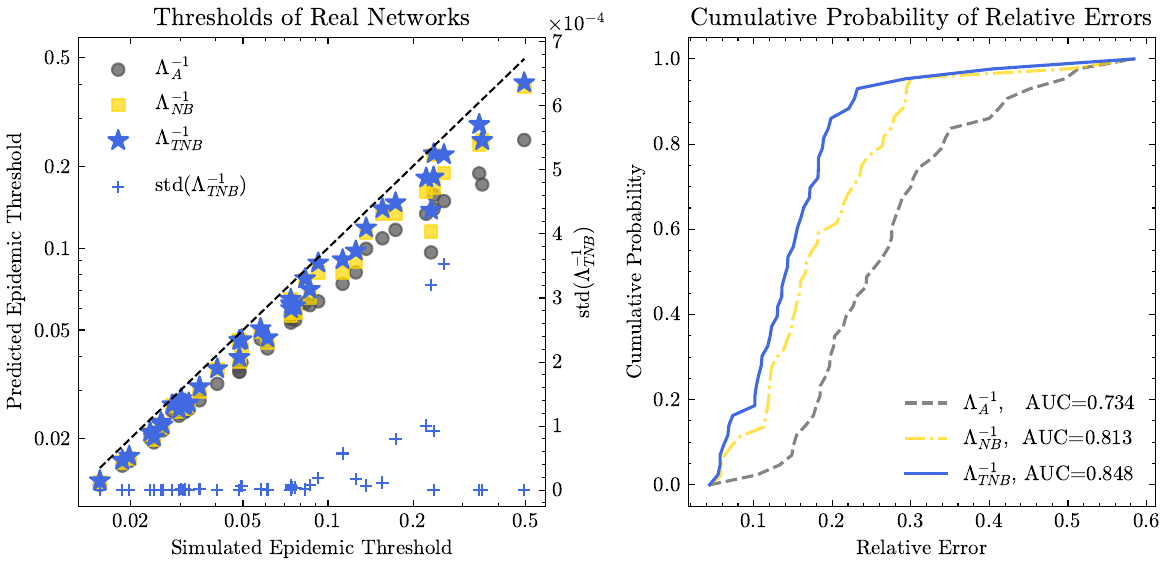}
    \caption{\textbf{Results of simulated and theoretical epidemic thresholds on 43 real-world networks:} (\textbf{\textit{left}}) Each marker represents the epidemic threshold of a real-world network as predicted by one of the three threshold predictions: $\Lambda_{A}^{-1}$, $\Lambda_{NB}^{-1}$, and $\Lambda_{TNB}^{-1}$. The simulated epidemic threshold is derived using the Gillespie method. The standard deviation of $\Lambda_{TNB}^{-1}$ is depicted on the right axis and is marked with the "+" symbol. (\textbf{\textit{right}}) The cumulative distribution of the relative error for each threshold prediction is presented. Additionally, the Area Under the Curve (AUC) for the three predictions are 0.734, 0.813, and 0.848, respectively.}
    \label{fig:real_nets_threshold}
\end{figure}

\paragraph{Real-world Networks}
To establish a benchmark for evaluating the predictive accuracy of those theoretical threshold predictions, we utilize a collection of 43 real-world networks, which is largely consistent with the dataset collection used in \cite{Wei_WWW_2017}. The dataset consist of a diverse range of networks, from social and infrastructural networks to animal and biological networks, each demonstrating distinct topological features. Their sizes range from 23 to 4941. All networks in our experiments are treated as undirected and unweighted, and only the largest connected component within each network is preserved. The source for all datasets is the Koblenz Network Collection \footnote{http://konect.unikoblenz.de/}.

\paragraph{Results}
In the left panel of Fig. \ref{fig:real_nets_threshold}, we report the estimated epidemic thresholds predicted by $\Lambda_{A}^{-1}$, $\Lambda_{NB}^{-1}$, and $\Lambda_{TNB}^{-1}$, as well as the thresholds produced by stochastic simulation. The results indicate that $\Lambda_{TNB}^{-1}$ consistently aligns closer to the simulated ones, underscoring the importance of accounting for local loops as in our theory. Moreover, we also report the standard deviation of $\Lambda_{TNB}^{-1}$, stemming from the randomly chosen operation in Eq.(\ref{equ:lrdmp_linear2}), computed from 10 independent trials. Among all the analyzed real-world networks, the standard deviations of $\Lambda_{TNB}^{-1}$ are less than $4\times 10^{-4}$, rendering it almost negligible.
To quantitatively assess the improvement achieved by $\Lambda_{TNB}^{-1}$, we define the relative error of the predicted threshold, $\hat{\lambda}_c$, in comparison to the simulated threshold, $\lambda_c$, as $(\lambda_c-\hat{\lambda}c)/\lambda_c$. In the right panel of Fig. \ref{fig:real_nets_threshold}, we present the cumulative probability distribution of the relative error for each theoretical method. The Area Under the Curve (AUC) serves as a metric to evaluate the overall performance of each method. Our results show that $\Lambda_{TNB}^{-1}$ offers improvements of 16\% and 4\% over the other two methods, respectively.

\begin{figure}
    \centering
    \includegraphics[width=0.8\textwidth]{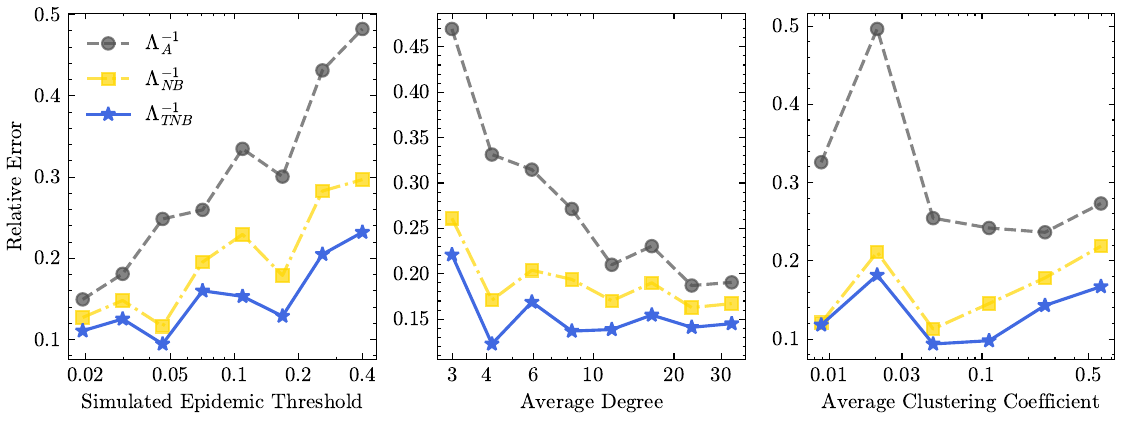}
    \caption{\textbf{Relative errors across 43 real-world networks:} (\textbf{\textit{left}}) based on the absolute simulated epidemic threshold; (\textbf{\textit{middle}}) corresponding to the average degree of the networks; and (\textbf{\textit{right}}) associated with the average clustering coefficient of the networks.}
    \label{fig:real_nets_threshold_2}
\end{figure}

In Fig. \ref{fig:real_nets_threshold_2}, we investigate the relative error of each method w.r.t. the values of simulated epidemic thresholds, average degree, and average clustering coefficient of real-world networks. We categorize the simulated epidemic thresholds into 8 bins and compute the average relative error for each bin. The results reveal that all methods generally provide better predictions when the "real" threshold is small, but their accuracy diminishes as the threshold increases. Particularly, $\Lambda_{TNB}^{-1}$ outperforms the other methods across all ranges, especially when the simulated threshold is large.
In a similar way, we divide the average degrees of the 43 real-world networks into 8 bins. The $\Lambda_{TNB}^{-1}$ displays a significant improvement over other methods in sparse networks (with lower average degrees). In denser cases, QMF has been proven to be a satisfactory approximation, meaning that our method can only achieve relatively small gains.
Further, to examine the impact of local loops on the relative errors, we categorize all networks into 7 bins based on their average clustering coefficient. As the results show, a larger clustering coefficient tends to result in a higher relative error. Still, $\Lambda_{TNB}^{-1}$ consistently improves upon existing methods in all cases. An odd observation is that all methods display anomalous failures when the clustering coefficient is around 0.02. For now, we attribute this counterintuitive result to outliers caused by our limited dataset.

\begin{figure}
    \centering
    \includegraphics[width=0.75\textwidth]{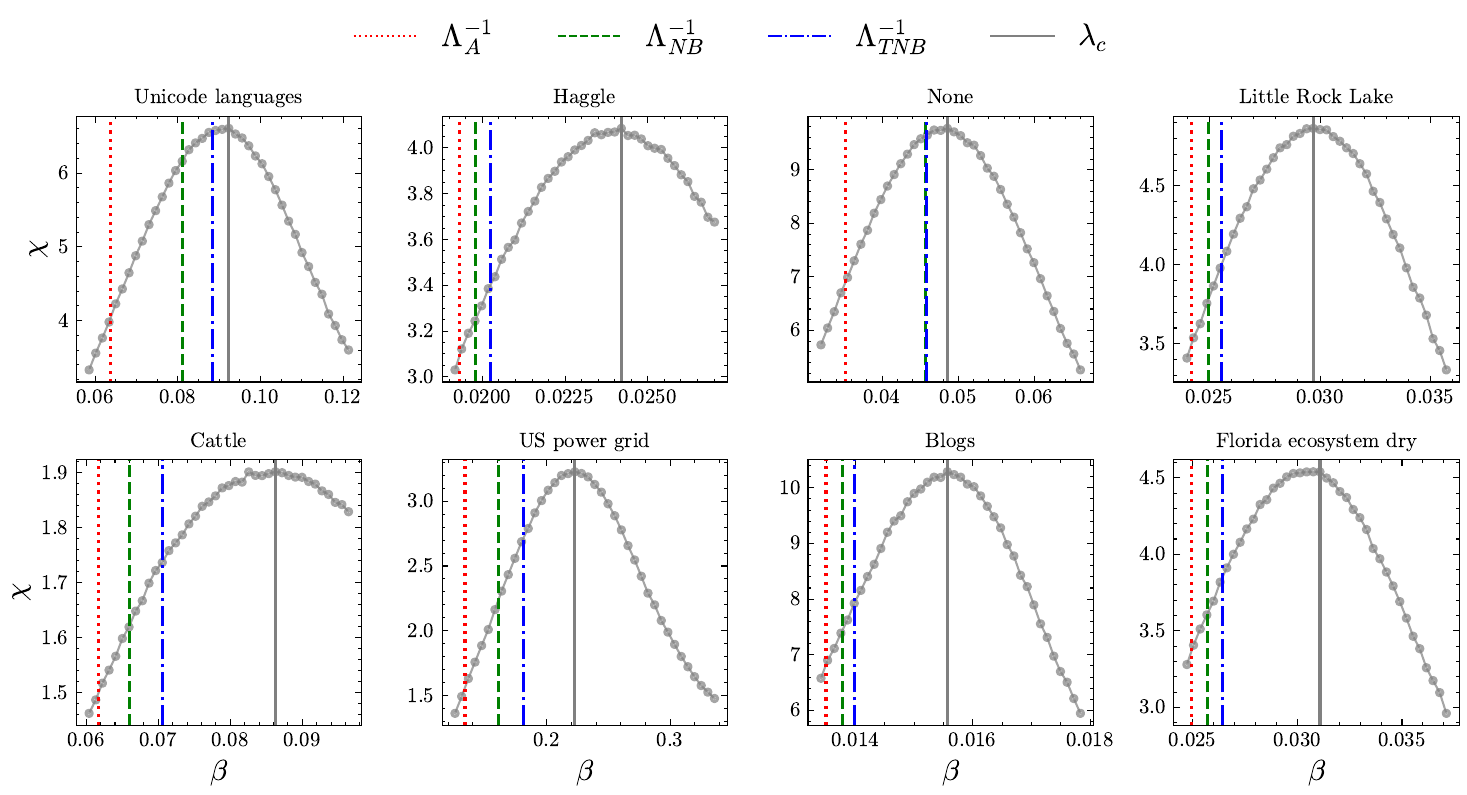}
    \caption{\textbf{Dynamical susceptibility curve for eight randomly selected real-world networks.}}
    \label{fig:real_nets_threshold_3}
\end{figure}

\paragraph{Visualization}
To visually probe the relationship between the theoretical and  simulated thresholds, we randomly select 8 real-world networks and plot their dynamical susceptibility curve $\chi_{\beta}$ as in Fig. \ref{fig:real_nets_threshold_3}, the peak of which is considered to be the indication of the epidemic threshold. We also plot three theoretical predictions alongside the curve. From the results in Fig. \ref{fig:real_nets_threshold_3}, there are three key observations: (1) The infection rate indicated by peak of $\chi_{\beta}$ serve as an upper bound to all three theoretical predictions; (2) Among the predictions, $\Lambda_{TNB}^{-1}$ offers the most accurate estimation; (3) In certain sparse networks, $\Lambda_{TNB}^{-1}$ reduces to $\Lambda_{NB}^{-1}$, i.e., predicts the same thresholds. These findings underscore the efficacy of $\Lambda_{TNB}^{-1}$ in providing a closer estimation to simulated thresholds, especially when compared to other methods. However, the relative improvements of $\Lambda_{TNB}^{-1}$ is limited, there is still a significant gap between the theoretical and simulated thresholds which demands further investigation in the future.

%% file: texs/sec_conclu.tex
\section{Conclusion and Discussion}

In this work, we aim to develop, in principle, a better approximation theory for recurrent-state epidemics in networks. Specifically, we correct and then extend the dynamic message-passing theory for recurrent-state epidemics to incorporate the local loops in the network, resulting in a loop-corrected recurrent dynamic message-passing method, or LrDMP. The proposed LrDMP prevents the echo chamber effect in local triangles by introducing new dynamic messages into directed triangles and allowing only non-backtracking propagation within edges and triangles. And we demonstrate that the linearized version of our method, utilizing the triangular non-backtracking (TNB) matrix , gives a new estimate of the epidemic threshold. In order not to overstate our results, we have clarified the theoretical relationship between LrDMP and other methods, as well as emphasized the limitation of LrDMP in cases where backtracking propagation is essential. Finally, empirical experiments on synthetic and 43 real-world networks show that the proposed LrDMP has better or competitive performance than other methods. Specifically, the LrDMP method achieves comparable accuracy in predicting epidemic prevalence to the pair approximation method, and offers improvements of 16\% and 4\% over the other two methods in the epidemic threshold estimation task, respectively.

The echo chamber effect occurs in all local loops, and incorporating triangles into message passing is the first attempt to reduce the self-amplification of network clustering to recurrent-state epidemics. In principle, we can generalize our method to consider higher-order dynamic messages to handle longer local loops. However, this will lead to more complex message-passing equations, making the approach less applicable when epidemics are sustained by backtracking propagation of hubs. It is interesting to develop a method for better performance by adaptively adjusting the length of the local loops it considers based on the clustering of the network. On the other hand, for unidirectional epidemics on networks, it is also necessary to rigorously develop loop-based message-passing methods, similar to those recently formulated for static problems \cite{newman2023message}. We will work on these challenges in future studies.

Our proposed method has the potential to benefit many network tasks, especially those that aim to minimize or maximize propagation prevalence by optimizing the initial network setup, including but not limited to influence maximization, immunization, and network robustness optimization. With its capability of providing quick and precise prevalence and threshold estimations, our method can efficiently prune the search space and speed up the search process of these tasks.

%% file: texs/dec.tex
\section{CRediT authorship contribution statement}
\textbf{Fei Gao}: Conceptualization, Formal analysis, Software, Validation, Visualization, Writing - original draft.
\textbf{Jing Liu}: Conceptualization, Investigation, Writing – review \& editing.
\textbf{Yaqian Zhao}: Funding acquisition, Project administration, Writing – review \& editing.

\section{Declaration of Competing Interest}

The authors declare that they have no known competing ﬁnancial interests or personal relationships that could have appeared to inﬂuence the work reported in this paper.

\section{Declaration of Generative AI and AI-assisted technologies in the writing process}

During the preparation of this work the authors used ChatGPT \footnote{https://chat.openai.com/} in order to facilitate the writing and debugging of experimental codes, and refine the manuscript's linguistic articulation.
After using this tool/service, the authors reviewed and edited the content as needed and takes full responsibility for the content of the publication.

%% file: main.bbl
\begin{thebibliography}{10}

\bibitem{PhysRevLett.111.068701}
Marian Bogu\~n\'a, Claudio Castellano, and Romualdo Pastor-Satorras.
\newblock Nature of the epidemic threshold for the susceptible-infected-susceptible dynamics in networks.
\newblock {\em Phys. Rev. Lett.}, 111:068701, Aug 2013.

\bibitem{burgio_network_2021}
Giulio Burgio, Alex Arenas, Sergio Gómez, and Joan~T. Matamalas.
\newblock Network clique cover approximation to analyze complex contagions through group interactions.
\newblock {\em Communications Physics}, 4(1):111, May 2021.

\bibitem{MPwL_2019}
George~T Cantwell and Mark~EJ Newman.
\newblock Message passing on networks with loops.
\newblock {\em Proceedings of the National Academy of Sciences}, 116(47):23398--23403, 2019.

\bibitem{Relevance_PRE_2018}
Claudio Castellano and Romualdo Pastor-Satorras.
\newblock Relevance of backtracking paths in recurrent-state epidemic spreading on networks.
\newblock {\em Phys. Rev. E}, 98:052313, Nov 2018.

\bibitem{PhysRevE.85.056111}
E.~Cator and P.~Van~Mieghem.
\newblock Second-order mean-field susceptible-infected-susceptible epidemic threshold.
\newblock {\em Phys. Rev. E}, 85:056111, May 2012.

\bibitem{QMF_PhysRevE.89.052802}
E.~Cator and P.~Van~Mieghem.
\newblock Nodal infection in markovian susceptible-infected-susceptible and susceptible-infected-removed epidemics on networks are non-negatively correlated.
\newblock {\em Phys. Rev. E}, 89:052802, May 2014.

\bibitem{10.1145/1284680.1284681}
Deepayan Chakrabarti, Yang Wang, Chenxi Wang, Jurij Leskovec, and Christos Faloutsos.
\newblock Epidemic thresholds in real networks.
\newblock {\em ACM Trans. Inf. Syst. Secur.}, 10(4), jan 2008.

\bibitem{IMP_2009}
Wei Chen, Yajun Wang, and Siyu Yang.
\newblock Efficient influence maximization in social networks.
\newblock In {\em ACM SIGKDD}, page 199–208, 2009.

\bibitem{QS_COSTA2021108046}
Guilherme~S. Costa and Silvio~C. Ferreira.
\newblock Simple quasistationary method for simulations of epidemic processes with localized states.
\newblock {\em Computer Physics Communications}, 267:108046, 2021.

\bibitem{Gillespie_COTA2017303}
Wesley Cota and Silvio~C. Ferreira.
\newblock Optimized gillespie algorithms for the simulation of markovian epidemic processes on large and heterogeneous networks.
\newblock {\em Computer Physics Communications}, 219:303--312, 2017.

\bibitem{Fundamentals-Exact}
Guilherme~Ferraz {de Arruda}, Francisco~A. Rodrigues, and Yamir Moreno.
\newblock Fundamentals of spreading processes in single and multilayer complex networks.
\newblock {\em Physics Reports}, 756:1--59, 2018.
\newblock Fundamentals of spreading processes in single and multilayer complex networks.

\bibitem{PhysRevE.86.041125}
Silvio~C. Ferreira, Claudio Castellano, and Romualdo Pastor-Satorras.
\newblock Epidemic thresholds of the susceptible-infected-susceptible model on networks: A comparison of numerical and theoretical results.
\newblock {\em Phys. Rev. E}, 86:041125, Oct 2012.

\bibitem{gao2022neural}
Fei Gao, Jiang Zhang, and Yan Zhang.
\newblock Neural enhanced dynamic message passing.
\newblock In {\em International Conference on Artificial Intelligence and Statistics}, pages 10471--10482. PMLR, 2022.

\bibitem{QMF_MMCA_2010}
S.~Gómez, A.~Arenas, J.~Borge-Holthoefer, S.~Meloni, and Y.~Moreno.
\newblock Discrete-time markov chain approach to contact-based disease spreading in complex networks.
\newblock {\em Europhysics Letters}, 89(3):38009, feb 2010.

\bibitem{HASHIMOTO1989211}
Ki~ichiro Hashimoto.
\newblock Zeta functions of finite graphs and representations of p-adic groups.
\newblock In K.~Hashimoto and Y.~Namikawa, editors, {\em Automorphic Forms and Geometry of Arithmetic Varieties}, volume~15 of {\em Advanced Studies in Pure Mathematics}, pages 211--280. Academic Press, 1989.

\bibitem{istvan2019mathematics}
Z~Istvan, KISS MILLER, C~SIMON JOEL, and L~Peter.
\newblock {\em Mathematics of Epidemics on Networks: From Exact to Approximate Models}.
\newblock Springer, 2019.

\bibitem{PhysRevE.82.016101}
Brian Karrer and M.~E.~J. Newman.
\newblock Message passing approach for general epidemic models.
\newblock {\em Phys. Rev. E}, 82:016101, Jul 2010.

\bibitem{BPwL_SA_2021}
Alec Kirkley, George~T Cantwell, and MEJ Newman.
\newblock Belief propagation for networks with loops.
\newblock {\em Science Advances}, 7(17):eabf1211, 2021.

\bibitem{NB_PNAS_2013}
Florent Krzakala, Cristopher Moore, Elchanan Mossel, Joe Neeman, Allan Sly, Lenka Zdeborová, and Pan Zhang.
\newblock Spectral redemption in clustering sparse networks.
\newblock {\em Proceedings of the National Academy of Sciences}, 110(52):20935--20940, 2013.

\bibitem{IMP_2018}
Yuchen Li, Ju~Fan, Yanhao Wang, and Kian-Lee Tan.
\newblock Influence maximization on social graphs: A survey.
\newblock {\em IEEE Transactions on Knowledge and Data Engineering}, 30(10):1852--1872, 2018.

\bibitem{Wei_WWW_2017}
Yuan Lin, Wei Chen, and Zhongzhi Zhang.
\newblock Assessing percolation threshold based on high-order non-backtracking matrices.
\newblock In {\em Proceedings of the 26th International Conference on World Wide Web}, WWW '17, page 223–232, Republic and Canton of Geneva, CHE, 2017. International World Wide Web Conferences Steering Committee.

\bibitem{DMP-SIR}
Andrey~Y. Lokhov, Marc M\'ezard, and Lenka Zdeborov\'a.
\newblock Dynamic message-passing equations for models with unidirectional dynamics.
\newblock {\em Phys. Rev. E}, 91:012811, Jan 2015.

\bibitem{Mata_2013}
Angélica~S. Mata and Silvio~C. Ferreira.
\newblock Pair quenched mean-field theory for the susceptible-infected-susceptible model on complex networks.
\newblock {\em Europhysics Letters}, 103(4):48003, sep 2013.

\bibitem{newman2018networks}
M.~Newman.
\newblock {\em Networks}.
\newblock OUP Oxford, 2018.

\bibitem{newman2003structure}
Mark~EJ Newman.
\newblock The structure and function of complex networks.
\newblock {\em SIAM review}, 45(2):167--256, 2003.

\bibitem{newman2023message}
MEJ Newman.
\newblock Message passing methods on complex networks.
\newblock {\em Proceedings of the Royal Society A}, 479(2270):20220774, 2023.

\bibitem{PhysRevE.105.024308}
Ernesto Ortega, David Machado, and Alejandro Lage-Castellanos.
\newblock Dynamics of epidemics from cavity master equations: Susceptible-infectious-susceptible models.
\newblock {\em Phys. Rev. E}, 105:024308, Feb 2022.

\bibitem{pastor2001epidemic}
Romualdo Pastor-Satorras and Alessandro Vespignani.
\newblock Epidemic spreading in scale-free networks.
\newblock {\em Physical review letters}, 86(14):3200, 2001.

\bibitem{pastor2002immunization}
Romualdo Pastor-Satorras and Alessandro Vespignani.
\newblock Immunization of complex networks.
\newblock {\em Physical review E}, 65(3):036104, 2002.

\bibitem{Fundamental_2022}
Daniel~J. Rosenkrantz, Anil Vullikanti, S.~S. Ravi, Richard~E. Stearns, Simon Levin, H.~Vincent Poor, and Madhav~V. Marathe.
\newblock Fundamental limitations on efficiently forecasting certain epidemic measures in network models.
\newblock {\em Proceedings of the National Academy of Sciences}, 119(4):e2109228119, 2022.

\bibitem{PhysRevE.94.042308}
Renan~S. Sander, Guilherme~S. Costa, and Silvio~C. Ferreira.
\newblock Sampling methods for the quasistationary regime of epidemic processes on regular and complex networks.
\newblock {\em Phys. Rev. E}, 94:042308, Oct 2016.

\bibitem{PhysRevE.89.022805}
Munik Shrestha and Cristopher Moore.
\newblock Message-passing approach for threshold models of behavior in networks.
\newblock {\em Phys. Rev. E}, 89:022805, Feb 2014.

\bibitem{Shrestha_rDMP}
Munik Shrestha, Samuel~V. Scarpino, and Cristopher Moore.
\newblock Message-passing approach for recurrent-state epidemic models on networks.
\newblock {\em Phys. Rev. E}, 92:022821, Aug 2015.

\bibitem{PhysRevResearch.1.033024}
Diogo~H. Silva, Silvio~C. Ferreira, Wesley Cota, Romualdo Pastor-Satorras, and Claudio Castellano.
\newblock Spectral properties and the accuracy of mean-field approaches for epidemics on correlated power-law networks.
\newblock {\em Phys. Rev. Res.}, 1:033024, Oct 2019.

\bibitem{PhysRevE.102.012313}
Diogo~H. Silva, Francisco~A. Rodrigues, and Silvio~C. Ferreira.
\newblock High prevalence regimes in the pair-quenched mean-field theory for the susceptible-infected-susceptible model on networks.
\newblock {\em Phys. Rev. E}, 102:012313, Jul 2020.

\bibitem{Comparison_PRE_2022}
Jos\'e Carlos~M. Silva, Diogo~H. Silva, Francisco~A. Rodrigues, and Silvio~C. Ferreira.
\newblock Comparison of theoretical approaches for epidemic processes with waning immunity in complex networks.
\newblock {\em Phys. Rev. E}, 106:034317, Sep 2022.

\bibitem{Nonbacktracking_2021}
Leo Torres, Kevin~S. Chan, Hanghang Tong, and Tina Eliassi-Rad.
\newblock Nonbacktracking eigenvalues under node removal: X-centrality and targeted immunization.
\newblock {\em SIAM Journal on Mathematics of Data Science}, 3(2):656--675, 2021.

\bibitem{4549746}
Piet Van~Mieghem, Jasmina Omic, and Robert Kooij.
\newblock Virus spread in networks.
\newblock {\em IEEE/ACM Transactions on Networking}, 17(1):1--14, 2009.

\bibitem{QMF_2009}
Piet Van~Mieghem, Jasmina Omic, and Robert Kooij.
\newblock Virus spread in networks.
\newblock {\em IEEE/ACM Trans. Netw.}, 17(1):1–14, feb 2009.

\bibitem{wang_predicting_2016}
Wei Wang, Quan-Hui Liu, Lin-Feng Zhong, Ming Tang, Hui Gao, and H.~Eugene Stanley.
\newblock Predicting the epidemic threshold of the susceptible-infected-recovered model.
\newblock {\em Scientific Reports}, 6(1):24676, April 2016.

\bibitem{Unification_Wang_2017}
Wei Wang, Ming Tang, H~Eugene Stanley, and Lidia~A Braunstein.
\newblock Unification of theoretical approaches for epidemic spreading on complex networks.
\newblock {\em Reports on Progress in Physics}, 80(3):036603, feb 2017.

\bibitem{1238052}
Yang Wang, D.~Chakrabarti, Chenxi Wang, and C.~Faloutsos.
\newblock Epidemic spreading in real networks: an eigenvalue viewpoint.
\newblock In {\em 22nd International Symposium on Reliable Distributed Systems, 2003. Proceedings.}, pages 25--34, 2003.

\bibitem{WS_clusterig}
Duncan~J Watts and Steven~H Strogatz.
\newblock Collective dynamics of ‘small-world’networks.
\newblock {\em nature}, 393(6684):440--442, 1998.

\bibitem{GBP_2000}
Jonathan~S Yedidia, William Freeman, and Yair Weiss.
\newblock Generalized belief propagation.
\newblock {\em Advances in neural information processing systems}, 13, 2000.

\bibitem{PhysRevE.96.042303}
Pan Zhang.
\newblock Spectral estimation of the percolation transition in clustered networks.
\newblock {\em Phys. Rev. E}, 96:042303, Oct 2017.

\end{thebibliography}
